\documentclass[aps,prc,reprint,groupedaddress,nofootinbib]{revtex4-1}
\pdfoutput=1 
\usepackage{graphicx}
\usepackage{color}
\usepackage[dvipsnames]{xcolor}
\usepackage{rotating}
\usepackage{bigints}
\usepackage{enumerate}

\usepackage[colorlinks=true,backref=false, linktocpage=true,
citecolor=blue,urlcolor=blue,linkcolor=blue,pdfpagemode=UseOutlines,pdfstartview=FitH,bookmarksnumbered=true,bookmarksopen]{hyperref}

\newcommand{\beq}{\begin{equation}}
\newcommand{\eeq}{\end{equation}}
\newcommand{\bea}{\begin{eqnarray}}
\newcommand{\eea}{\end{eqnarray}}

\newcommand{\calO}{{\cal O}}

\newcommand{\MeV}{\, {\rm MeV}}

\newenvironment{paolo}{\bf \color{MidnightBlue}}{}
\newcommand{\bepa}{\begin{paolo}}
\newcommand{\enpa}{\end{paolo}}

\begin{document}
\title{Quartic cumulant of baryon number in the presence of QCD critical point}

\author{D. Mroczek}
\affiliation{Illinois Center for Advanced Studies of the Universe, Department of Physics, University of Illinois at Urbana-Champaign, Urbana, IL 61801, USA}

\author{A. R. Nava Acuna}
\affiliation{Department of Physics, University of Houston, Houston, TX, USA  77204}

\author{J. Noronha-Hostler}
\affiliation{Illinois Center for Advanced Studies of the Universe, Department of Physics, University of Illinois at Urbana-Champaign, Urbana, IL 61801, USA}

\author{P. Parotto}
\affiliation{University of Wuppertal, Department of Physics, Wuppertal D-42119, Germany}

\author{C. Ratti}
\affiliation{Department of Physics, University of Houston, Houston, TX, USA  77204}

\author{M.A. Stephanov}
\affiliation{Physics Department, University of Illinois at Chicago, Chicago, IL 60607, USA}

%https://www.overleaf.com/project/5d417804d53e373773532de8
\date{\today}
\
\begin{abstract}
  In the context of the ongoing search for the QCD critical point at
  the Relativistic Heavy-Ion Collider, we study the equation of state
  near the critical point in the temperature and baryon chemical
  potential plane. We use the parametric representation introduced in
  earlier literature, which maps the universal 3D Ising equation of
  state onto the QCD phase diagram using several non-universal
  parameters. We focus on the quartic cumulant of the baryon number,
  or baryon number susceptibility~$\chi_4^B$, which can be accessed
  experimentally via net-proton fluctuation kurtosis measurements.  It
  was originally predicted, through universality arguments based on
  the {\em leading} singular contribution, that $\chi_4^B$ and
  net-proton kurtosis should show a specific non-monotonic behavior
  due to the critical point. In particular, when following the
   freeze-out curve on the phase diagram by decreasing beam
  energy, the kurtosis is expected to dip, and then peak, when the
  beam energy scan passes close to the critical point.  We
  study the effects of the non-universal and thus far unknown
  parameters of the Ising-to-QCD mapping on the behavior
  of~$\chi_4^B$.  We find that, while the peak remains a solid
  feature, the presence of the critical point does not necessarily
  cause a dip in $\chi_4^B$ on the freezeout line {\em below} the
  transition temperature. The critical point contribution to the dip
  appears only for a narrow set of mapping parameters, when subleading
  singular terms are sufficiently suppressed.
\end{abstract}
%\keywords{}

\maketitle

\section{Introduction}

One of the current major thrusts of the nuclear physics program is to
map out the phase diagram of Quantum Chromodynamics (QCD) and
specifically look for a critical point in the transition from a hadron
resonance gas into deconfined plasma of quarks and gluons. Because the
location of the QCD critical point is yet unknown, searches are
currently ongoing across the relevant region of the QCD phase
diagram. At high temperatures and intermediate baryon chemical
potentials, relativistic heavy-ion collisions are able to scan the
phase diagram by systematically decreasing the collision energy. This
is the motivation behind the second phase of the Beam Energy Scan
(BES-II) at Relativistic Heavy-Ion Collider (RHIC) (see, e.g.,
Ref.~\cite{Bzdak:2019pkr} for a recent review). At lower temperatures and
higher baryon chemical potentials, useful information can be extracted
from the study of neutron stars and neutron star mergers. In fact, it
appears that there may even be significant overlap in the phase
diagram pertaining to the lowest beam energies in heavy-ion collisions
and neutron star mergers
\cite{Most:2018eaw,Adamczewski-Musch:2020slf}.

Lattice QCD calculations cannot be performed at finite~$\mu_B$
\cite{Troyer:2004ge}; therefore, it is currently not possible to
determine the location of the critical point from first
principles. Thus, experimental searches for the critical point are
central to determining its location
\cite{Adam:2020unf,Adamczewski-Musch:2020slf}.  The main strategy is
based on the search for certain non-monotonic dependence of
fluctuations on an experimental variable, such as the collision energy
$\sqrt s$, as the critical region is traversed during the scan of the
QCD phase diagram
\cite{Stephanov:1998dy,Stephanov:1999zu,Stephanov:2008qz,Stephanov:2011pb}. The
nonmonotonic behavior of fluctuation measures is directly related to
the divergence of susceptibilities at the critical point.  Therefore,
susceptibilities of conserved charges are of major interest for first
principle lattice calculations (see, e.g., Ref.~\cite{Ratti:2018ksb}
for a review).  In the case of heavy-ion collisions, there are three
conserved charges: baryon number ($B$), strangeness ($S$), and
electric charge ($Q$), whereas in neutron star (mergers) only $B$ and
$Q$ are conserved, because the typical time scales are sufficiently
large for weak processes to become relevant.

Baryon number susceptibilities diverge at the critical point
\cite{Hatta:2002sj}, and are, therefore, the most promising observables
in its search. Since experiments measure multiplicities of {\em
  charged} particles, the closest quantities to baryon number
susceptibilities, or cumulants, are the net-proton number cumulants,
which show similar critical
behavior~\cite{Hatta:2003wn,Stephanov:2008qz,Athanasiou:2010kw}. Electric
charge fluctuations contain a singular contribution from net-proton
fluctuations, but this effect is diluted by pions and therefore it is
expected to be milder \cite{Hatta:2003wn}.  Additionally, {\em higher
  order} cumulants are the most sensitive to critical behavior because
they scale with higher powers of the correlation length
\cite{Stephanov:2008qz,Stephanov:2011pb} in the vicinity of the
critical point. However, experimental measurements currently are only
available up to the fourth
cumulant~\cite{Aggarwal:2010wy,Adam:2020unf} at large baryon densities
with reasonable error bars.\footnote{The data for the sixth cumulant \cite{Nonaka:2020crv}
  are also available but with large statistical error bars and only at
  vanishing baryon densities.}

At $\mu_B=0$, it is possible to calculate the higher order $BSQ$ susceptibilities 
on the lattice and then use them to reconstruct the lower order ones at small 
finite baryon densities, although with large numerical uncertainties 
\cite{Borsanyi:2018grb,Bazavov:2017tot,Ratti:2018ksb}. Alternatively, effective 
models exist that can reproduce lattice QCD results and do include a critical 
point at finite baryon density \cite{Critelli:2017oub}. 

Another approach is to make use of the fact that the QCD critical
point is expected to be in the same universality class as the 3D-Ising
model
\cite{Rajagopal:1992qz,Berges:1998rc,Halasz:1998qr,Karsch:2001nf,deForcrand:2003vyj}.
Using this approach, a specific non-monotonic behavior of the
fourth cumulant of net-proton number as a function of $\sqrt{s}$
was proposed as a potential critical point signature in
Ref.~\cite{Stephanov:2011pb}. This prediction has sparked interest in
the community, especially in light of the BES-II and its Fixed
Target Program \cite{STARnote,Cebra:2014sxa}, which is intended to
provide larger statistics and reach lower collision energies. 

The
baryon number susceptibility, which has a similar behavior, can be
obtained from the equation of state by differentiating the pressure at
fixed temperature:
\begin{equation}
\chi_4^B(T,\mu_B)=\left(\frac{\partial^4p}{\partial\mu_B^4}\right)_T\,.
\label{eq:chi4B}
\end{equation}
Due to the mapping between the QCD and the 3D Ising model critical
equations of state, the leading divergence at the critical point comes from
the fourth derivative of the Gibbs free energy~$G$, i.e., the third derivative of the critical
order parameter (the magnetization $M$) with
respect to the ordering (magnetic) field $h$ at constant reduced temperature~$r$:
\begin{equation}\label{eq:kappa4}
    \chi_4^{\rm Ising} (r,h) 
= \left( \frac{\partial^4 G}{\partial h^4} \right)_r 
= \left( \frac{\partial^3 M}{\partial h^3} \right)_r \, \, .
\end{equation}

Taking only the leading singular contribution, the predicted behavior
for $\chi_4^B$ along a freeze-out curve (location of freezeout point
as a function of $\sqrt s$) starting at
$\mu_B=0$ and passing close to the critical point is as follows. From
its value at $\mu_B=0$, $\chi_4^B$ is expected to decrease
at increasing $\mu_B$, then move upwards and reach a peak in the
vicinity of the critical point.  This peculiar, doubly non-monotonic behavior
has motivated the experimental search for the critical point in the past
years, also due to a quite similar behavior observed in the measured quantity
\begin{equation}
\kappa \sigma^2 = \kappa_4/\kappa_2\,,
\label{eq:ks2}
\end{equation}
where $\kappa$, $\sigma=\sqrt{\kappa_2}$ and $\kappa_4$ are the kurtosis,
variance and quartic cumulant of the 
net-proton number distribution. Indeed, the
data from the STAR experiment \cite{Aggarwal:2010wy} show
$\kappa \sigma^2$ decreasing and then swinging upwards as the
collision energy decreases, which resembles the behavior predicted in
Ref.~\cite{Stephanov:2011pb}. Although this similarity is indeed quite
promising, other explanations have been proposed for the dip, such as
global conservation of baryon number -- which is expected to play a
bigger role at low collision energies where the system is smaller
\cite{Braun-Munzinger:2016yjz,Braun-Munzinger:2020jbk}. Transport
models that do not include any criticality, but do account for charge conservation, are
able to reproduce the decrease at finite $\mu_B$
\cite{Sahoo:2012wn}. On the other hand, the dip also
arises when extrapolating $\chi_4^B$  to finite
$\mu_B$ in lattice QCD through a Taylor series
\cite{Bazavov:2017tot,Borsanyi:2018grb}. This suggests that at least some contribution
to the experimentally observed dip comes from the equilibrium equation of state, which may, in principle, be due to the approach to the critical point.

The specific non-monotonic behavior predicted in
Ref.~\cite{Stephanov:2011pb} and described above focuses on the leading
contribution to $\chi_4^B$, given by $\chi_4^{\rm Ising}$. In the
parametric equation of state we use in this paper, due to the mixing
of $r$ and $h$ variables in the mapping of 3D Ising to QCD equation of
state, there are also subleading critical contributions. The
peculiarity of the QCD equation of state, as we see below in more
detail, is that the leading contribution is suppressed by the
smallness of the slope $\alpha_1$ of the phase-separating line in the
$T,\mu_B$ plane at the critical point. Therefore, unless the $r,h$ mixing
is also suppressed, the subleading critical contribution could
dominate in a significant part of the critical region, thus
qualitatively changing the prediction. 

In this work we investigate this effect by comparing two choices of
the mixing parameters, which show qualitatively different behavior of
$\chi_4^B$ near the critical point. One choice is a common ``default''
choice in the literature, where the $r,h$ mixing is not
suppressed. Another choice is motivated by the recent work in
Ref.\cite{Pradeep:2019ccv}, which argues that close to the chiral
(small quark mass) limit, the mixing is suppressed. While in the
latter choice we recover the pattern of $\chi_4^B$ behavior similar to
Ref.~\cite{Stephanov:2011pb}, in the former, the subleading terms
significantly changes that pattern. While the peak of $\chi_4^B$ is a
robust feature independent of the parameter choice, the dip at
$\mu_B<\mu_{BC}$ is sensitive to the choice. It is worth pointing out that we explored several other parameter choices, not
shown here, and that the dip disappears in almost all of them. The second parameter choice shown here is one of the few
in which the dip is still visible. The reason for this will become clear below.

This paper is organized as follows. In Section~\ref{sec:Ising_to_QCD}
we quickly summarize the procedure developed in
Ref.~\cite{Parotto:2018pwx} to construct equations of state for QCD
with a built-in criticality in the correct universality class.
In Section~\ref{sec:size-critical-region} we present a discussion of
the dependence of the critical region size and shape on the
different parameters, focusing on the contribution from the leading
divergence. \ In Section~\ref{sec:results} we present our results
for several different choices of the parameters in the Ising-to-QCD
map, which lead to our conclusions, summarized in
Section~\ref{sec:concl}.

\section{Parametric equation of state}\label{sec:Ising_to_QCD}
\label{map}

In this work, we utilize the procedure for constructing a family of
equations of states with a critical point developed in
Ref.~\cite{Parotto:2018pwx}. This parametric family  is
constructed in such a way that all its members match lattice QCD results at
$\mu_B=0$ (up to order $\calO (\mu_B^4)$) and contain a critical point
in the 3D Ising model universality class. We note that the
implementation of the critical behavior is essentially the same as in
Ref.~\cite{Stephanov:2011pb}.

The procedure can be summarized as follows:
\begin{enumerate}[\bf i.]
\item Define a parametrization of the 3D Ising model EoS in the vicinity of the 
critical point, imposing the correct critical behavior. Express the magnetization 
$M$, the magnetic field $h$ and the reduced temperature $r = (T - T_c) / T_c$ 
in terms of the new parameters $(R,\theta)$ with \cite{Nonaka:2004pg,Guida:1996ep,Schofield:1969zz,Bluhm:2006av}: 
\begin{align} \label{eq:param_Ising} \nonumber
M &= M_0 R^\beta \theta \, \, , \\
h &= h_0 R^{\beta \delta} \tilde{h}(\theta) \, \, , \\ \nonumber 
r &= R (1 - \theta^2) \, \, ,
\end{align}
where $M_0 \simeq 0.605$ and $h_0 \simeq 0.364$ are normalization 
constants, $\tilde{h} (\theta) = \theta (1 + a \theta^2 + b \theta^4)$, with 
$a= - 0.76201$ and $b=0.00804$, and 
$\beta \simeq 0.326$, $\delta \simeq 4.80$ are 3D Ising model critical 
exponents \cite{Guida:1996ep}. The parameters are within the range 
$R \geq 0$ and $\left| \theta \right| \leq \theta_0$, where 
$\theta_0 \simeq 1.154$ is the first nontrivial zero of $\tilde{h} (\theta)$.
\item Map the phase diagram of the 3D-Ising model onto the $T\mu_B$
  plane of QCD, 
\textit{choosing} the location of the critical point. A simple linear map \cite{Rehr:1973zz} requires 
six parameters, and can be written as:
\begin{align}
\frac{T - T_C}{T_C} &=  w \left( r \rho \,  \sin \alpha_1  + h \, \sin \alpha_2 \right) \, \, , 
\label{eq:IsQCDmap1} \\ 
\frac{\mu_B - \mu_{BC}}{T_C} &=  w \left( - r \rho \, \cos \alpha_1 - h \, \cos \alpha_2 \right) \, \, , 
\label{eq:IsQCDmap2}
\end{align} 
where $(T_C,\mu_{BC})$ are the coordinates of the critical point, and 
$(\alpha_1, \alpha_2)$ are the angles between the horizontal 
(fixed $T$) lines on the QCD phase diagram and the $h=0$ and $r=0$ 
Ising model axes, respectively. Finally, $w$ and $\rho$ are scaling parameters for the 
Ising-to-QCD map: $w$ determines the overall scale of both $r$ and $h$, 
while $\rho$ determines the relative scale between the two.

As  in Ref.~\cite{Parotto:2018pwx}, we reduce the number of 
parameters to four by imposing that the critical point is located on the chiral 
transition line given by lattice QCD calculations \cite{Bellwied:2015rza}:
\begin{equation}\label{eq:trline}
T = T_0 + \kappa_2 \, T_0 \left( \frac{\mu_B}{T_0} \right)^2 + {\cal O} (\mu_B^4),
\end{equation}
which allows us to fix the values of $T_C$ and $\alpha_1$ by choosing 
$\mu_{BC}$ only.

In order to be consistent with previous work, we use the same input from 
lattice QCD as in Ref.~\cite{Parotto:2018pwx}. Although recently new 
results on the QCD transition line have become available\footnote{Both in 
this work and in Ref~\cite{Parotto:2018pwx}, we assume that the QCD 
transition line is a parabola, with curvature $\kappa_2$ determined in 
Ref.~\cite{Bellwied:2015rza}. Recent results from lattice 
QCD~\cite{Bazavov:2018mes,Borsanyi:2020fev} are consistent with this value 
of the curvature, and predict the next to leading order parameter $\kappa_4$ 
which is consistent with 0 within error-bars.} 
\cite{Bazavov:2018mes,Borsanyi:2020fev}, we note that utilizing these new
results would not have any effect on the conclusions presented here.

\item Impose exact matching to lattice QCD at $\mu_B=0$ at the level of 
the coefficients of Taylor expansion of the pressure through:
\begin{equation} \label{eq:coeffs}
T^4 c_n^{\text{LAT}} (T) = T^4 c_n^{\text{Non-Ising}} (T) + T_C^4 c_n^{\text{Ising}} (T) \, \, ,
\end{equation}
where $c_n^{\text{LAT}}$ are the coefficients calculated from the
lattice, and $c_n^{\text{Ising}}$ determine the contribution to the
former due to the presence of the critical
point. Eq.~(\ref{eq:coeffs}) is thus the definition for the coefficients
$c_n^{\text{Non-Ising}}$ required to match the given critical equation of
state to lattice data without changing the singular behavior at the
critical point.  The procedure is carried out up to order
$\calO (\mu_B^4)$.
\item Reconstruct the full QCD pressure as:
\begin{multline} \label{eq:Pfull}
P (T, \mu_B) = T^4 \sum_n c_n^{\text{Non-Ising}} (T) \left( \frac{\mu_B}{T} \right)^n \\+ 
P^{\text{QCD}}_{\text{crit}}(T, \mu_B) \, \, ,
\end{multline}
where $P^{\text{QCD}}_{\text{crit}}(T, \mu_B)$ is the critical pressure from the 3D-Ising model mapped 
onto QCD. For additional details, we again refer the 
reader to Ref.~\cite{Parotto:2018pwx}.
\end{enumerate}

With the procedure summarized here, the constructed EoS (i.e. the
pressure, from which all needed derivatives can be calculated) by
construction meets the initial requirements, and depends on the
non-universal mapping between 3D-Ising model and QCD through the
specific choice of parameters.

In the following we will consider only the critical
point contribution to the 4-th order susceptibility of the baryon
number $\chi_4^B$ in Eq.~(\ref{eq:chi4B}).

Since the procedure we just summarized 
stops at order $\calO (\mu_B^4)$, the total contribution obtained in our 
approach  differs from the critical one by a constant in $\mu_B$, i.e. a function 
depending on the temperature only. Thus, a similar plot for the total 
contribution would show the same features.

\section{The size and shape of the critical region}
\label{sec:size-critical-region}

While the divergence of $\chi_4^B$ at the critical point is present
for any choice of parameters due to the parametrization in
Eq.~(\ref{eq:param_Ising}), the extent of the region in the phase
diagram where its magnitude is large (either positive or negative) is
a nonuniversal property of the theory -- the ``size of the
critical region'' -- which cannot be inferred from universality
arguments. It is nonetheless of crucial importance, as it can
ultimately determine whether the critical behavior can be observed in
experiments.

Here we describe how the parameters of the mapping control the size of
the critical region. We define the critical region as the region where
the leading singular part of the equation of state dominates over the
regular part. This comparison cannot be done on the pressure itself,
since the critical contribution to the pressure vanishes at the
critical point (as $r^{2-\alpha}$). A reasonable measure of the
critical region should be based on a quantity which {\em diverges} at the
critical point, such as the baryon susceptibility,
$\chi_2^B=P_{\mu\mu}$ or, in our case, $\chi_4^B=P_{\mu\mu\mu\mu}$
(where $P_\mu=\partial (p/T^4)/\partial(\mu_B/T)$ at fixed $T$). We
shall estimate the size of the critical region along the
crossover, $h=0$, line. The singular part of $\chi_4^B$ at $h=0$ is
given by
\begin{eqnarray}
\label{eqn5}
\chi_4^{\rm sing} &\sim& AG_{\mu\mu\mu\mu}(r,0)\sim AG_{hhhh}(r,0) h_{\mu}^4 \\
&\sim& Ar^{\beta(1-3\delta)}\left(\frac{s_1}{wT_C s_{12}}\right)^4 \\
&\sim& A\left(\frac{\Delta \mu_B}{\rho w T_C c_1}\right)^{\beta(1-3\delta)}
\left(\frac{s_1}{wT_C s_{12}}\right)^4 \, \, .
\nonumber
\end{eqnarray}
where $P_{crit}^{QCD}(T,\mu_B) = AG(r,h)$, $G_\mu=\partial G/\partial(\mu_B/T)$, $s_i=\sin\alpha_i,\, c_i=\cos\alpha_i$ and
$s_{12}=\sin(\alpha_1-\alpha_2)$, $A$ is an overall constant and 
$h_\mu=\partial h/\partial \mu_B$ at fixed~$T$. Comparing this to the regular
contribution of order $\chi_4^{\rm reg}\sim 1$, we find for the extent of
the critical region in the $\mu_B$ direction:
\begin{equation}
\Delta \mu_{\rm B}\sim T_C\rho w c_1
\left(\frac{A^{1/4}}{ T_C}\frac{s_1 }{w s_{12}}\right)^{\frac{4}{\beta(3\delta-1)}}.
\label{eq:DmuB}
\end{equation}

Therefore, while increasing $\rho$ increases the
size of the critical region, the effect of increasing the parameter
$w$ is very weak. For the mean-field value of $\beta=1/2$ and
$\delta=3$, the $w$ dependence is completely absent, while for the
values $\beta=1/3$, $\delta=5$ approximating the exact values of 3D Ising
model exponents one finds a very weak dependence 
$\Delta\mu_{\rm B} \sim w^{1/7}$.

To determine the extent in the vertical, i.e.
$\mu_B= {\rm const} = \mu_{BC}$ direction, we note that this 
corresponds to a finite ratio $h/r=-\rho c_1/c_2$. Thus, the scaling variable 
$r/h^{1/(\beta\delta)}\to0$ as we approach the critical point, and we can set 
$r=0$ when determining the magnitude of $\chi_4$:
\begin{eqnarray}
\label{chi4,r=0}
\chi_4^{\rm sing} &\sim& A G_{\mu\mu\mu\mu}(0,h)\sim AG_{hhhh}(0,h) h_{\mu}^4 \\
&\sim& Ah^{(1-3\delta)/\delta}\left(\frac{s_1}{wT_C s_{12}}\right)^4 \\
&=& A\left(\frac{c_1\Delta T}{w T_C s_{12}}\right)^{(1-3\delta)/\delta}
\left(\frac{s_1}{wT_C s_{12}}\right)^4 \, \, .
\nonumber
\end{eqnarray}

The condition $\chi_4^{\rm sing} \sim 1$ then gives
\begin{equation}
\Delta T\sim T_C 
\left(\frac{A}{ T_C^4}\right)^{\frac{\delta}{3\delta-1}}
\frac{s_1}{c_1}
\left(\frac{s_1 }{w s_{12}}\right)^{\frac{\delta+1}{3\delta-1}} \, \, .
\label{eq:DT}
\end{equation}
The dependence on $w$ is given by $\Delta T\sim
w^{-\frac{\delta+1}{3\delta-1}}$. For the mean-field value of
$\delta$ this corresponds to $w^{-1/2}$ and for $\delta=5$ to $w^{-3/7}$.

\section{Results and discussion}\label{sec:results}

We now employ the procedure described in Section \ref{map} to calculate the 
susceptibilities of the baryon number. We summarize our parameter choices in 
Table~\ref{tab:param_choices}. We fix the location of the critical
point sufficiently far from the $\mu_B=0$ axis to allow for maximum
freedom in our parameter choice but  still within the range of the
Taylor expansion of $\mathcal{O}(\mu_B^4)$.  To satisfy those criteria we use 
$\mu_{BC} = 420 \MeV$, which results in $T_C \simeq 138 \MeV$ and 
$\alpha_1 \simeq 4.6 ^\circ$, and study several values of the parameters 
$(w,\rho)$. In addition, we consider two different choices for the relative angle 
between the $(r,h)$ axes.  First, we keep the two axes orthogonal 
($\alpha_2 - \alpha_1 = 90 ^\circ$), as this has been a common
``default'' choice in the literature. Then we examine the case with the angle 
between the two axes $\alpha_2 - \alpha_1 = -3 ^\circ$. This second choice is 
motivated by the fact that, in the chiral limit, the angle
difference vanishes (as quark mass to power $2/5$) and
$0<\alpha_2<\alpha_1$ for sufficiently small quark mass~\footnote{This
  can be seen explicitly in the Random Matrix Model of the QCD phase diagram~\cite{Halasz:1998qr,Pradeep:2019ccv}.}, according to
Ref.~\cite{Pradeep:2019ccv}. 
Note that, according to  Eqs.~\eqref{eq:DmuB} and~\eqref{eq:DT}, a small 
value for $s_{12}$ yields a larger critical region size for the
same $w$ and~$\rho$: $\Delta \mu_{B} \sim s_{12}^{-6/7}$ 
and $\Delta T \sim s_{12}^{-3/7}$.

\begin{table}
\begin{tabular}{| c | c | c | c | c | c | c |}
 \hline
& $\mu_{BC}$ & $T_C$ & $\alpha_1$ & $\alpha_2 - \alpha_1$ & $w$ & $\rho$ \\
 \hline
{\bf I.} & $420 \MeV$ & $138 \MeV$ & $4.6^\circ$ & $90^\circ$ & $0.5, 1, 2$ & $0.5, 1, 2 $ \\
  \hline
{\bf II.} & $420 \MeV$ & $138 \MeV$ & $4.6^\circ$ & {\bf $-3^\circ$} & $0.5, 1, 2$ & $0.5, 1, 2$ \\
\hline
\end{tabular}
\caption{The two sets of parameter choices we employ in this work. Notice 
that, as detailed in the main text, $T_C$ and $\alpha_1$ are not free 
parameters, but they follow from the choice of $\mu_{BC}$ due to the 
constraints from Eq. (\ref{eq:trline}).}
\label{tab:param_choices}
\end{table}

\begin{figure*}
    \centering
    \includegraphics[width=\textwidth]{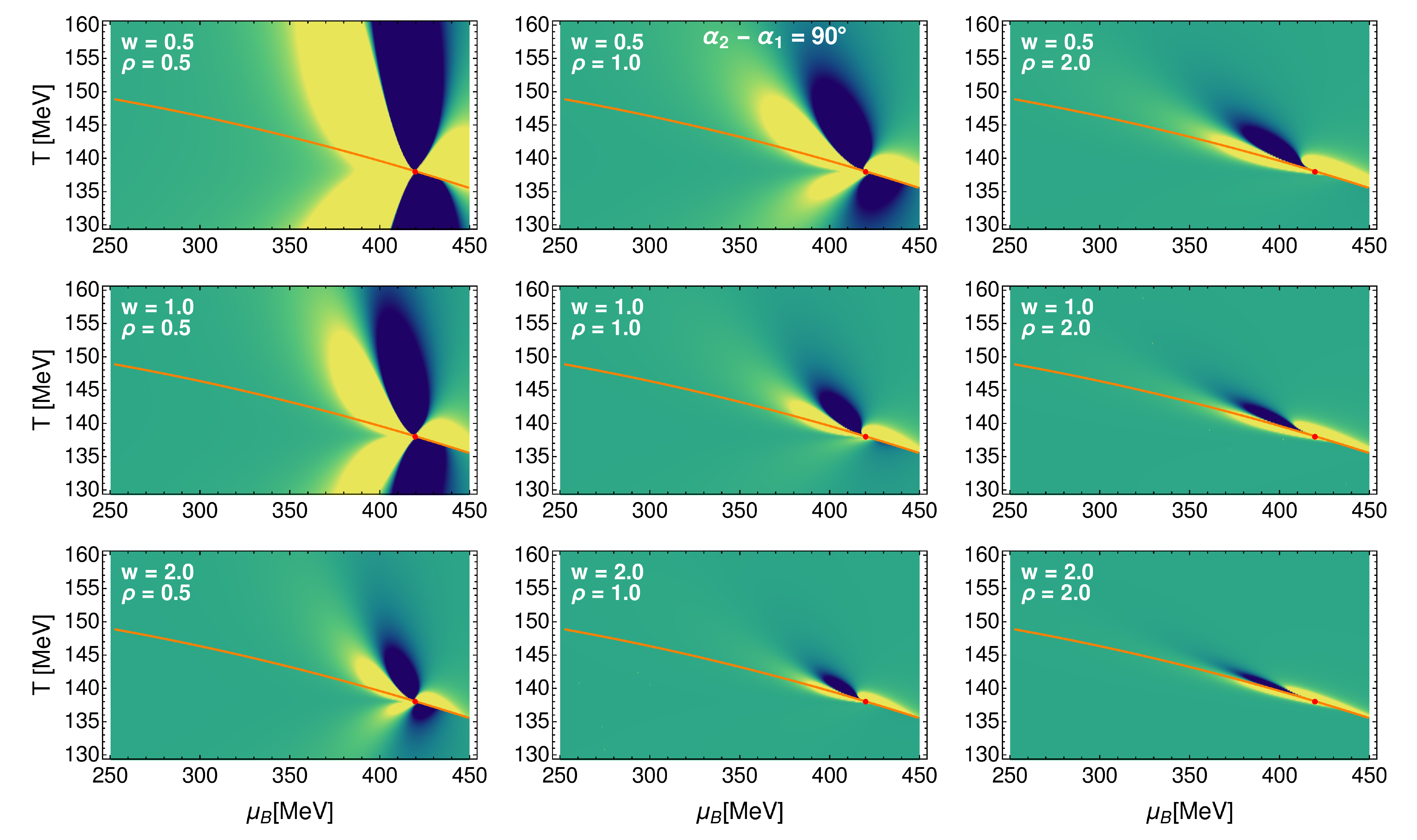}
    \caption{Density plots of the critical contribution to $\chi_4^B(T,\mu_B)$ in 		
    the $(T,\mu_B)$ plane with a critical point located at 
    $(T_C \simeq 138 \MeV, \mu_{BC} = 420\MeV)$, and with 
    $\alpha_2 - \alpha_1 = 90 ^\circ$, for (top to bottom) $w = 0.5,~ 1, ~2$ and (left 
    to right) $\rho = 0.5, ~1, ~2$. The critical point is indicated by a red dot, while 
    the chiral/deconfinement transition line is represented by the solid orange 
    line. The yellow and green areas correspond to positive values (the 
    regions where it is the largest are indicated in yellow) of $\chi_4^B$, while 
    the blue ones correspond to negative values (darker blue in the regions 
    where it is largest in magnitude).}
    \label{fig:chi4_contour_angle90}
\end{figure*}

\begin{figure*}
    \centering
    \includegraphics[width=\textwidth]{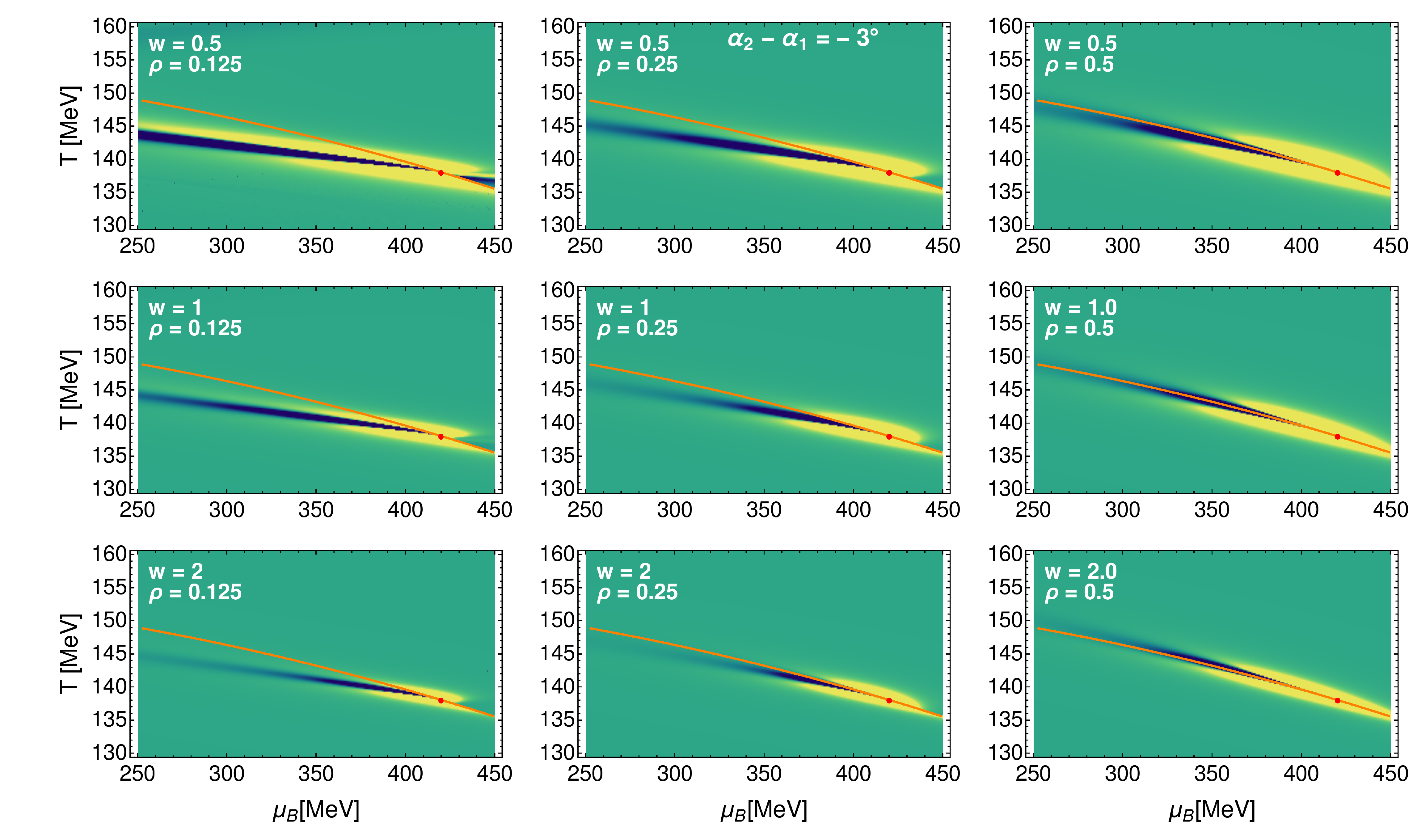}
    \caption{Density plots of the critical contribution to $\chi_4^B(T,\mu_B)$ in 
    the $(T,\mu_B)$ plane with a critical point located at 
    $(T_C \simeq 138 \MeV, \mu_{BC} = 420\MeV)$, and with 
    $\alpha_2 - \alpha_1 = -3 ^\circ$, for (top to bottom) $w = 0.5,~ 1,~ 2$ and (left 
    to right) $\rho = 0.125,~ 0.25,~ 0.5$. The critical point is indicated by a red dot, while 
    the chiral/deconfinement transition line is represented by the solid, orange 
    line. The yellow and green areas correspond to positive values (the 
    regions where it is the largest are indicated in yellow) of $\chi_4^B$, while 
    the blue ones correspond to negative values (darker blue in the regions 
    where it is largest in magnitude).}
    \label{fig:chi4_contour_anglem3}
\end{figure*}

We now investigate the behavior of the critical contribution to
$\chi_4^B$
over the QCD 
phase diagram, with focus on the region close to the critical point 
$T = 130 - 160 \MeV$ and $\mu_B = 250 - 450 \MeV$.

In Figs.~\ref{fig:chi4_contour_angle90} 
and \ref{fig:chi4_contour_anglem3}, density plots of the critical 
contribution to $\chi_4^B(T,\mu_B)$ in 
the $(T,~\mu_B)$ plane are shown for $w = 0.5, 1, 2$ and $\rho = 0.5, 1, 2$ in 
the case of $\alpha_2 - \alpha_1 = 90 ^\circ$ and $w = 0.5, 1, 2$ and 
$\rho = 0.125, 0.25, 0.5$ in the case of $\alpha_2 - \alpha_1 = -3 ^\circ$, 
respectively. The yellow and green areas 
correspond to positive values (the regions where it is the largest are 
indicated in yellow) of $\chi_4^B$, while the blue ones correspond to negative 
values (darker blue in the regions where it is largest in magnitude). 
The orange curve shows the QCD 
transition line from Eq. (\ref{eq:trline}). The red dot marks the
critical point.

We note that the color function is not the same for
Figs.~\ref{fig:chi4_contour_angle90} and \ref{fig:chi4_contour_anglem3}. The 
color schemes are such that a factor 10 in the value of $\chi_4^B$ separates 
the two figures, for the same color. This is because, due to the 
dependence of $\chi_4^B$ on $s_{12}$, this quantity is overall significantly larger in 
all the plots of Fig.~\ref{fig:chi4_contour_anglem3} than in those of 
Fig.~\ref{fig:chi4_contour_angle90}.

We would like to point out the following relevant features in
Figs.~\ref{fig:chi4_contour_angle90} and \ref{fig:chi4_contour_anglem3}:
\begin{enumerate}[\bf i.]
\item A smaller value of $w$ leads to a larger critical region in the $T$ direction, for both values of the relative 
angle $\alpha_2 - \alpha_1$. This follows from Eq.(\ref{eq:DT});

\item The main effect of $\rho$ is to stretch the critical region in the $\mu_B$ direction. Indeed, the size of the critical
region along $\mu_B$ increases linearly with $\rho$, while the one in the $T$ direction is not
affected by $\rho$ according to Eqs.~(\ref{eq:DmuB}) and~(\ref{eq:DT}).

\item  
It is most interesting to compare our findings to what was originally
anticipated in Ref.~\cite{Stephanov:2011pb} based on the leading
singular contribution. While the pattern in Fig.~\ref{fig:chi4_contour_anglem3} is in agreement with
the leading singularity prediction, in Fig.~\ref{fig:chi4_contour_angle90} that prediction only
holds extremely close to the critical point. 

Away from the critical point the subleading singular terms modify the
pattern. In Fig.~\ref{fig:chi4_contour_angle90} for $\rho=2.0$ and in
Fig.~\ref{fig:chi4_contour_anglem3} the main effect is the bending of
the negative lobe away from the crossover line. The downward bending
in Fig.~\ref{fig:chi4_contour_anglem3} is a consequence of
$0<\alpha_2<\alpha_1$, while the upward bending in
Fig.~\ref{fig:chi4_contour_angle90} is a consequence of
$\alpha_1<\alpha_2<180^\circ$, as explained in
Ref.\cite{Pradeep:2019ccv}.  

As a result, in
Fig.~\ref{fig:chi4_contour_angle90}, the critical contribution to the
{\em dip} to the left of the critical point is absent, except in the
extremely close vicinity of the critical point. Instead, the approach to the
critical point from the left is characterized by a peak instead of a
dip. Furthermore, for smaller $\rho$ values, an additional negative lobe
appears below the critical point for {\em larger} $\mu_B$.

\end{enumerate}

To understand the effect of the choice of $\alpha_2$ on the
significance of the subleading singular contributions to~$\chi_4^B$ we
observe, let us examine the Ising-to-QCD mapping more closely.
Eqs.~(\ref{eq:IsQCDmap1}),~(\ref{eq:IsQCDmap2}), allow us to convert
the derivatives with respect to $\mu_B$ in the definition of
$\chi_4^B$ in Eq.~(\ref{eq:chi4B}) into derivatives with respect to Ising variables $h$ and
$r$:
\begin{align}
    \partial_{\mu_B} 
  &=\frac{1}{w \, \rho \, T_C s_{12}} 
    \left( s_1 \, \partial_h + s_2 \, \partial_r\right)
 \,.
\end{align}
Since $h$ corresponds to the most relevant perturbation at the
critical point ($h$ has the largest scaling dimension), the dominant
contribution to the derivative $\partial_{\mu_B}$ sufficiently close
to the critical point comes from $\partial_h$.  Since~$\alpha_1$ is
small, when $\alpha_2$ is {\em not} small, the contribution of $\partial_h$
is suppressed by $s_1/s_2$ compared to $\partial_r$. This is precisely
the case in Fig.~\ref{fig:chi4_contour_angle90}.  While taking only
the most divergent terms corresponds to setting
$\partial_{\mu_B} \sim \partial_h$, and hence
$\chi_4^B \sim \chi_4^{\rm Ising}$ from Eq.~(\ref{eq:kappa4}), the
full expression for $\chi_4^B$ contains many additional subleading,
less singular terms which involve $\partial_r$.  The subleading terms
will become negligible sufficiently close to the critical point, but
if the leading contribution is strongly suppressed this may not happen
until we are extremely close to the critical point, as seen in
Fig.~\ref{fig:chi4_contour_angle90}. Thus, the pattern of the $T\mu_B$
dependence of $\chi_B$ around the critical point is significantly
affected by the subleading terms in this scenario.

On the other hand, when $\alpha_2$ {\em is} small, as for our choice
$\alpha_2\approx 1.6^\circ$, the pattern is indeed more similar to the
one described in Ref.~\cite{Stephanov:2011pb}. This can be seen in
Fig.~\ref{fig:chi4_contour_anglem3}, especially when $\rho = 0.5$.

\begin{figure*}
\includegraphics[width=\textwidth]{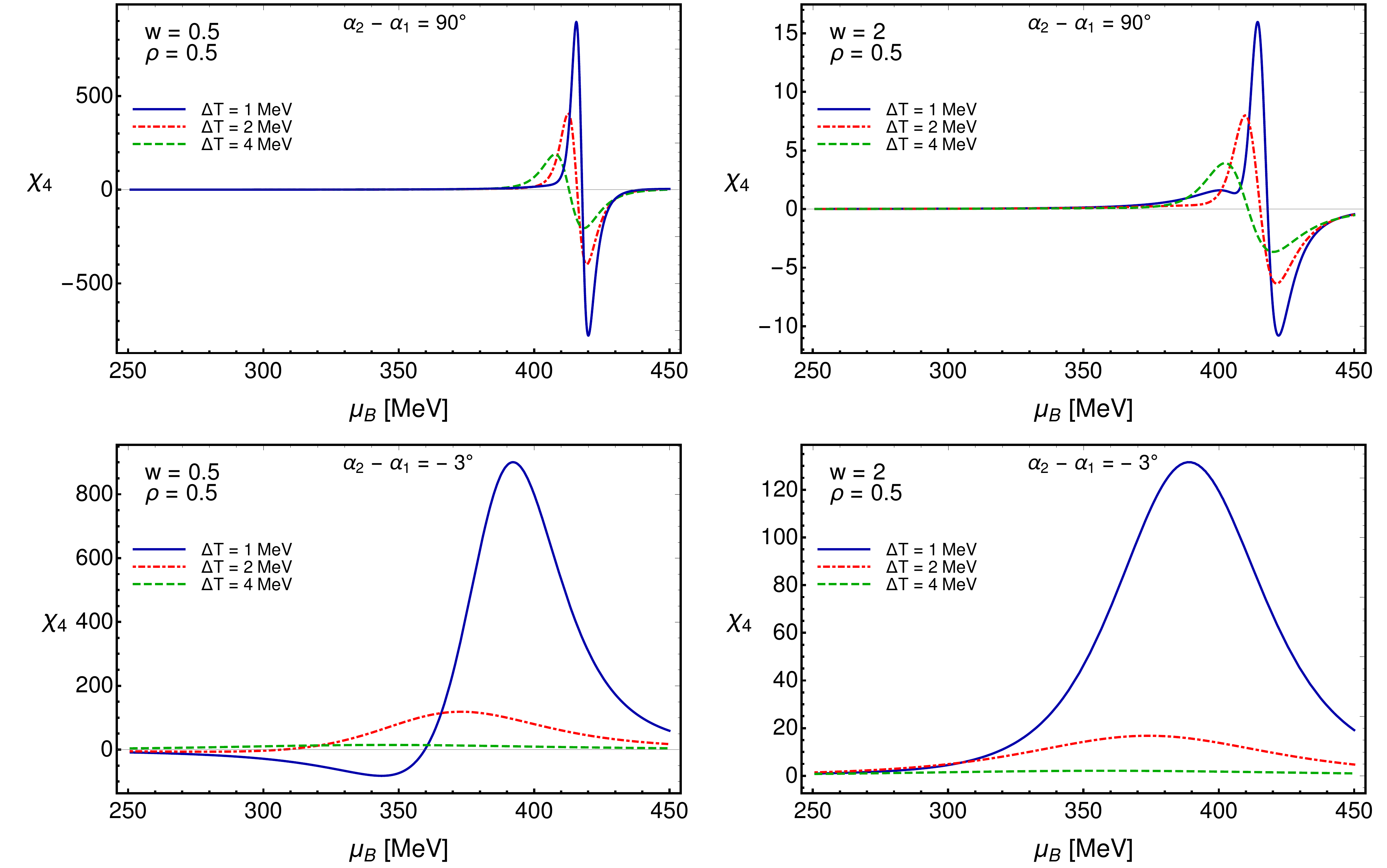}
\caption{Profile of the critical contribution to $\chi_4^B$ along lines parallel to the chiral transition line, and separated by $\Delta T = 1,2,4$. The top and bottom rows correspond to $\alpha_2-\alpha_1 = 90 ^\circ$ and $\alpha_2-\alpha_1 = -3 ^\circ$, respectively. }
\label{fig:deltaTplot}
\end{figure*}

After analyzing the general behavior of $\chi_4^B$ over the QCD phase 
diagram, we now wish to determine the impact that its features can have on 
experimental measurements. We shall make a simplifying assumption that
net-proton kurtosis has a similar critical behavior to $\chi_4^B$, following
the argument of Ref.~\cite{Hatta:2003wn}.
In the following we study the behavior of $\chi_4^B$ along exemplary 
freeze-out trajectories, which are roughly parallel to the chiral/deconfinement transition 
line from Eq.~\eqref{eq:trline}: 
\begin{equation}\label{eq:trline2}
    T_{\rm F}(\mu_B) = T_0 + \kappa_2 \, T_0 \left( \frac{\mu_B}{T_0} \right)^2 - \Delta T \, \, ,
\end{equation}
where $\Delta T$ indicates the shift in temperature downward from the
transition line.  In Fig.~\ref{fig:deltaTplot} we show the behavior of
the critical contribution to $\chi_4^B$ along such lines, with shifts
$\Delta T = 1,2,4 \MeV$. In the different panels, we consider the
cases with $\alpha_2 - \alpha_1 = 90 ^\circ$ (top row) and
$\alpha_2 - \alpha_1 = - 3 ^\circ$ (bottom row), and with the
parameter choices $w = \rho = 0.5$ (left column) and
$w = 2, \rho = 0.5$ (right column).

\begin{figure}
\includegraphics[width=\linewidth]{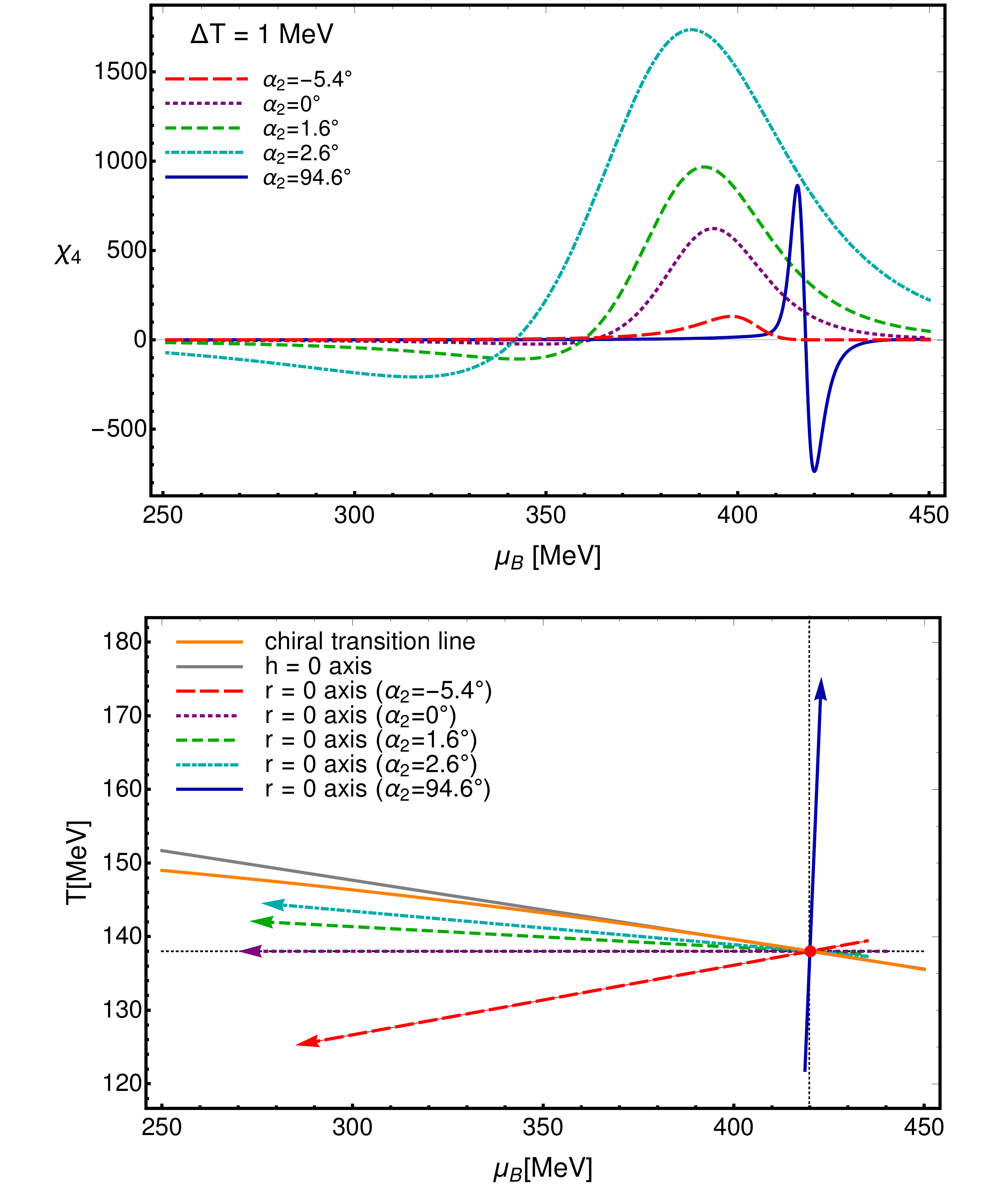}
\caption{(Top panel) Profile of the critical contribution to $\chi_4^B$ along 
lines parallel to the chiral transition line, and separated by $\Delta T = 1$. The 
different lines correspond to different choices for 
$\alpha_2 = -5.4 ^\circ, 0 ^\circ, 1.6 ^\circ, 2.6 ^\circ, 94.6 ^\circ$. For all these 
curves we fixed $w = \rho = 0.5$.(Bottom panel): The chiral transition line 
(orange) is shown together with the $h=0$ axis (gray) and the $r=0$ axis 
corresponding to the choices shown in the top panel. The color coding is kept 
the same. }
\label{fig:anglesplot}
\end{figure}

The choice that displays a dip for $\mu_B<\mu_{BC}$ is the
one with $w = \rho = 0.5$, $\alpha_2 - \alpha_1 = -3 ^\circ$ and only
in the close vicinity of the transition line, i.e., for
$\Delta T = 1, 2 \MeV$.  Fig.~\ref{fig:chi4_contour_anglem3} suggests
that this would be the case also for smaller values of $\rho$, as we
note that the lower the value of $\rho$, the more apparent the
downward bending is of the negative (blue) lobe. Since this behavior
follows from our choice for the angle $\alpha_2$, we consider
in the top panel of Fig.~\ref{fig:anglesplot} different choices for
the angle $\alpha_2$. We focus on lines parallel to the transition
line, with $\Delta T = 1 \MeV$, and keep $w = \rho = 0.5$ in all
cases. 

We consider a handful of choices for the angle $\alpha_2$. We include
the ones corresponding to Fig.~\ref{fig:chi4_contour_angle90}
($\alpha_2 \simeq 94.6 ^\circ$) and
Fig.~\ref{fig:chi4_contour_anglem3} ($\alpha_2 \simeq 1.6 ^\circ$), as
well as $\alpha_2 \simeq - 5.4 ^\circ, 0 ^\circ, 2.6 ^\circ$. In the
bottom panel of Fig.~\ref{fig:anglesplot} we show the orientations of
the $r=0$ axis corresponding to the different values of $\alpha_2$ we
used. As anticipated, only in the cases satisfying
$0 < \alpha_2 < \alpha_1$ a dip for $\mu_B<\mu_{BC}$ is
seen. Moreover, we consider in this plot a shift $\Delta T = 1 \MeV$
between the chemical freeze-out line and the chiral transition
line. With larger separation, a dip would be harder to observe, as
shown in Fig.~\ref{fig:deltaTplot}.

\section{Experimental considerations}\label{sec:exp}

In our current study we focused on the {\em equilibrium} properties of
the QCD equation of state that can lead to the potential discovery of
the QCD critical point.  However, because heavy-ion collisions are
inherently dynamical systems, direct comparison with experimental data
would require an event-by-event relativistic viscous hydrodynamics
model with BSQ conserved charges \cite{Noronha-Hostler:2019ayj,Monnai:2019hkn} and critical fluctuations coupled to
a hadronic transport code.

While important efforts are being
made along these lines in terms of new hydrodynamical models
\cite{Karpenko:2013wva,Karpenko:2013ama,Du:2019obx,Denicol:2018wdp,Batyuk:2017sku,Fotakis:2019nbq,Dore:2020fiq},
transport coefficients
\cite{Demir:2008tr,Denicol:2013nua,Kadam:2014cua,Monnai:2016kud,Rougemont:2017tlu,Rougemont:2017tlu,Auvinen:2017fjw,Martinez:2019bsn},
critical fluctuations
\cite{Stephanov:1999zu,Jiang:2015hri,Stephanov:2017ghc,Nahrgang:2018afz,An:2019osr},
and freeze-out
\cite{Stephanov:2009ra,Feng:2016ddr,Li:2017via,Steinheimer:2019iso,Oliinychenko:2019zfk,Oliinychenko:2020cmr},
the full dynamical description does not yet exist at this time.  In
the meantime, a number of attempts have been made to quantify effects
such as critical slowing down and memory, finite volume/lifetime, number of particles, decays, charge
conservation, kinematic cuts, low statistics etc
\cite{Berdnikov:1999ph,Mukherjee:2015swa,Mukherjee:2016kyu,Akamatsu:2018vjr,Westfall:2014fwa,Noronha-Hostler:2016rpd,Bzdak:2017ltv,Steinheimer:2017dpb,Bzdak:2016jxo,Bzdak:2018axe,Braun-Munzinger:2019yxj,Poberezhnyuk:2020ayn,Braun-Munzinger:2020jbk}. Yet
further studies have looked into the influence of far-from-equilibrium
initial conditions and potential attractors at the critical point
\cite{Dore:2020jye} and the influence of viscous effects across a
first order phase transition line \cite{Feng:2018anl}.

Another remaining question that is very relevant to this study is the temperature difference between hadronization and freeze-out.   
Earlier attempts were made in dynamic models to quantify either the time scale or temperature range in the difference between hadronization and freeze-out \cite{Wong:1996va,Rapp:2000gy,Greiner:2000tu,Greiner:2004vm,NoronhaHostler:2007jf,NoronhaHostler:2009cf,Noronha-Hostler:2014aia,Almasi:2014pya,Beitel:2016ghw,Gallmeister:2017ths}.  Generally, this depends on the number of hadrons in the system \cite{Alba:2020jir} and their corresponding interactions \cite{Becattini:2012xb,Alba:2016hwx,Vovchenko:2017zpj,Lo:2017lym}. However, given enough particles that appear near the phase transition that are strongly interacting, it is possible to reach chemical equilibrium on very short time scales \cite{Borsanyi:2013hza,Bellwied:2013cta,Borsanyi:2014ewa,Alba:2014eba,Bellwied:2018tkc}.

\section{Conclusions}\label{sec:concl}

In this work we have studied the fourth order susceptibility,
$\chi_4^B$, of the baryon number in QCD in the presence of a critical
point in the 3D Ising model universality class. We found that some
features of the $T-$ and $\mu_B-$dependence of $\chi_4^B$ could be
significantly affected by sub-leading, less singular terms in the
critical behavior.  In all cases that we studied, we found a diverging
peak at the critical point. However, only in the special case of
$0<\alpha_2<\alpha_1$ (which also implies a wide critical region that
is extended along the chiral phase transition) do we obtain a dip as
one approaches the critical point along an exemplary freezeout curve {\em
  below} the transition temperature. In this case, at temperatures
significantly lower than the transition the dip moves to smaller
$\mu_B$ and fades away.

One of the conclusions which can be drawn from this study is
that the peak in net-proton kurtosis is a more robust signature of the critical
point than the dip. However, it is also important to keep in mind that the
observation of the dip may help determine or constrain the value of
the parameter $\alpha_2$, provided other potential experimental
contributions to the dip (the baseline) are under control.

It is important to emphasize that this study only considers the
equilibrium equation of state and it would be interesting and
important to explore these issues further in {\em dynamical} models. For
example, as has been observed in
Refs.\cite{Berdnikov:1999ph,Stephanov:2009ra,Mukherjee:2016kyu},
critical slowing down, charge conservation and memory effects may help
to preserve the signatures of critical fluctuations down to lower
temperatures below the critical region.

\section*{Acknowledgments}
This material is based upon work supported by the National Science
Foundation under grant no. PHY-1654219 and by the U.S. Department of
Energy, Office of Science, Office of Nuclear Physics, within the
framework of the Beam Energy Scan Topical (BEST) Collaboration and
grants Nos. DE-SC0019175 and DE-FG02-01ER41195.  We also acknowledge
the support from the Center of Advanced Computing and Data Systems at
the University of Houston. J.N.H. acknowledges support from the Alfred
P Sloan Foundation. P.P. also acknowledges support by the DFG grant
SFB/TR55. D.M. was supported by the National Science Foundation
Graduate Research Fellowship Program under Grant No. DGE – 1746047

\bibliography{reference}

\end{document}